\documentclass[11pt]{article}
 \usepackage{color}
\usepackage{amsmath}
\usepackage{amssymb}
\usepackage{amsthm}
\usepackage{epstopdf}
\usepackage{graphicx}
\usepackage{mathtools}
\usepackage[usenames,dvipsnames, table]{xcolor}
\usepackage[hang]{caption}
\usepackage{hyperref}
\usepackage{oldgerm}  
\usepackage{float}
\usepackage{marginnote}

\newcommand{\nco}{\newcommand}
\nco{\one}{\ensuremath{\,\,\mathrm{l}\!\!\!1}} 
\nco{\ZZ}{\mathbb{Z}}
\nco{\CC}{\mathbb{C}}

\nco{\yellow}{\color{yellow}}
\nco{\green}{\color{green}}
\nco{\red}{\color{red}}
\nco{\cyan}{\color{cyan}}
\nco{\blue}{\color{blue}}
\nco{\violett}{\color{violet}}
\nco{\magenta}{\color{magenta}}
\nco{\redend}{\normalcolor}
\nco{\0}{\underline{0}}

\definecolor{violet}{rgb}{1,0,1}
 
\def\ie{{\rm i.e.,\/}\ }

\def\be{\begin{equation}}\def\ee{\end{equation}}
\def\bea{\begin{eqnarray}}\def\eea{\end{eqnarray}}
\def\bee{\begin{enumerate}}\def\eee{\end{enumerate}}
\def\bei{\begin{itemize}}\def\eei{\end{itemize}}

\def\ommit#1{{}}

 \def\CC{{\cal C}}

 \def\eq=#1{\buildrel #1 \over{=}}

 \def\cm{\checkmark}
 \def\gg{\mathfrak{g}}
 \def\bmu{\bar\mu}
  \def\bm{\bar{m}}
   \def\bm{\overline{m}}
 
 \nco{\rnc}{\renewcommand}
\rnc{\title}[1]{{\Large\bf\mbox{}\\\medskip#1\bigskip\medskip\\}}
\rnc{\author}[1]{{\large #1\smallskip\\}}
\nco{\address}[1]{{\em #1\medskip\\}}

\begin{document}
\begin{titlepage}
\begin{center}
\title{
{Conjugation Properties of Tensor Product and Fusion Coefficients}}
\medskip
\author{Robert Coquereaux} 
\address{Aix Marseille Univ, Univ Toulon, CNRS, CPT, Marseille, France}
\author{Jean-Bernard Zuber}
\address{Laboratoire de Physique Th\'eorique et des Hautes \'Energies, CNRS UMR 7589, \\
Universit\'e Pierre et Marie Curie, Sorbonne Universit\'es, 4 Place Jussieu, 75252 Paris Cedex 05
 }
 \bigskip\medskip
\today
\bigskip\medskip

\begin{abstract}
\noindent {We review some recent results on properties of tensor product and fusion coefficients
under complex conjugation of one of the factors.
Some of these results have been proven, some others are conjectures awaiting 
a proof, one of them involving hitherto unnoticed observations on
ordinary representation theory of finite simple groups of Lie type. }
\end{abstract}
\end{center}


\vspace*{70mm}
\end{titlepage}

%%%%%%%%%%%%%%%%%%%%%%%%%%%%%%%%%%%%

\section{Notations and Results}

\subsection{Notations}
\label{notations}
In the following $\lambda$, $\mu$, etc label either %inequivalent 
finite dimensional irreps of a simple Lie algebra $\gg$
or of the corresponding simply connected compact Lie group $G$; 
or (for the sake of comparison) irreps of a finite group $\Gamma$; 
or %(inequivalent) 
integrable irreps of an affine algebra $\widehat{\gg}_k$ at a finite integral level $k$.  
\\
Following a standard abuse of notations, for the Lie groups and algebras, $\lambda$ denotes both 
the highest weight of the representation and the representation itself.

$N_{\lambda\mu}^\nu$ denotes respectively 
the coefficients of decomposition of the tensor product $\lambda\otimes \mu$ into inequivalent  irreps $\nu$
(Littlewood--Richardson coefficients) of $G$ or $\Gamma$; 
or the coefficients of decomposition of the fusion product denoted $\lambda\star \mu$ into irreps $\nu$ of $\widehat{\gg}_k$.

It is often convenient to regard this set of coefficients as  elements of matrices, thus
\be N_{\lambda\mu}^\nu=(N_\lambda)_\mu^\nu \,.\ee

These coefficients satisfy the sum rule 
\be \label{dimsumrule}\dim_\lambda \dim_\mu=\sum_\nu N_{\lambda\mu}^\nu \dim_\nu\ee
where $\dim_\alpha$ denotes the dimension, resp. the {\it quantum dimension}, 
of the irrep $\alpha$ of $G$, $\gg$ or $\Gamma$,
resp. of $\widehat{\gg}$.
When $\lambda$ refers to a representation of complex type, we denote by $\bar\lambda$ the 
(equivalence class of its) complex conjugate.
Recall that among the simple Lie algebras, only those of type $A_r$, any $r$; $D_{r}$, $r$ odd; and $E_6$  admit 
complex representations. 

For a given pair $(\lambda,\mu)$, consider the moments of the $N$'s
$$ m_r := \sum_\nu (N_{\lambda\mu}^\nu)^r \qquad\qquad r\in\mathbb{N}\,.$$
In particular, $m_0$ counts the number of distinct (\ie non equivalent) $\nu$'s appearing in 
 the decomposition of $\lambda\otimes\mu$, resp. $\lambda\star \mu$. 
 
For non real $\lambda$ and $\mu$, we want to compare $m_r$ and 
$\bm_r:=\sum_\nu (N_{\lambda\bmu}^\nu)^r $.

Call $\mathfrak P$ 
the property that the {\it multisets} $\{N_{\lambda\mu}^\nu\}$ and $\{N_{\lambda\bmu}^{\nu'}\}$ are 
identical. Since for given $\lambda$ and $\mu$
these multisets are finite, there is an equivalence
\be m_r =\bm_r\quad \forall r\in \mathbb{N} \qquad \Leftrightarrow\qquad \mathfrak P \,.\ee

\subsection{A list of results and open questions}
We  start with a fairly obvious statement\\[8pt]
{\bf Proposition 1} \cite{RCJBZ2014}{\sl\  For any Lie group, any finite group or any affine Lie algebra,
$m_2=\bm_2$, \ie $\sum_\nu (N_{\lambda\mu}^\nu)^2=\sum_{\nu'} (N_{\lambda\bmu}^{\nu'})^2$,}
\\
see below the (easy) proof in sect.\  1.3. Much more surprising is the following

{\bf Proposition 2} \cite{RCJBZ2011}{\sl\  For any simple Lie algebra,  or any affine simple  Lie algebra, 
$m_1=\bm_1$, \ie $\sum_\nu N_{\lambda\mu}^\nu=\sum_{\nu'} N_{\lambda\bmu}^{\nu'} $.
This is not generally true  for finite groups $\Gamma$.
This is not generally true for quantum doubles of finite groups either} \cite{RCJBZ2013}.

{\bf Problem 1} {\sl For a given finite group $\Gamma$, find a criterion on $(\Gamma, \lambda,\mu)$ for Prop.\  2 to hold.}

{\bf Proposition 3} \cite{RCJBZ2014}{\sl\  For the Lie group SU(3),  
$m_r=\bm_r$ for all $r$, \ie we have property $\mathfrak P$. 
Moreover we know a (non-canonical and non-unique)
piece-wise linear bijection $(\nu,\alpha)\leftrightarrow (\nu',\alpha')$, where $\alpha$ is a multiplicity index running over
$N_{\lambda\mu}^\nu$ values. 
This property $\mathfrak P$  is not true in general for higher rank
 SU($N$) nor for other Lie groups .}

{\bf Proposition 4} \cite{RCJBZ2014}{\sl\  For the affine algebra $\widehat{su}(3)$ at finite level $k$, 
$m_r=\bm_r$ for all $r$, \ie we have property $\mathfrak P$. 
This is not true in general for higher rank $\widehat{su}(N)$ or other affine algebras.}

This is, however, satisfied by low-level representations.\\
{\bf Problem 2} {\sl For each $\widehat{\gg}_k$, find a criterion on $(\lambda,\mu,k)$ for Prop.\  4 to hold.}

Also missing in  $\widehat{su}(3)$ is a general mapping $\nu\leftrightarrow \nu'$  compatible with 
the level. Although we found one in a few particular cases, a general expression is still missing.\\
{\bf Problem 3} {\sl For each level in $\widehat{su(3)}_k$, find a 
piece-wise linear bijection $\nu\leftrightarrow \nu'$}.

A weaker property than property $\mathfrak P$, which 
follows from it, is that $m_0=\bm_0$. 

{\bf Proposition 5} \cite{RCJBZ2014}{\sl\  For the affine algebra $\widehat{su}(3)$ at finite level $k$,
$m_0=\bm_0$. This seems to be also true for $\widehat{su}(4)$, but 
this is not true in general for higher rank $\widehat{su}(N)$, $N\ge 5$,  or other affine algebras.}

This is, however, satisfied by low-level representations.\\
{\bf Problem 4}  {\sl For each $\widehat{\gg}_k$, find a criterion on $(\lambda,\mu,k)$ for Prop.\  5 to hold.}

These results on the equality of various $m_k$ and $\bar m_k$ are summarized in Table \ref{results-ext}.

\bigskip 
\begin{table}[h]
\begin{center}
$
\begin{array}{|c||c||c||c||c||c||c|} 
\hline
{\mbox{ }} & SU(3)  \ {\rm or }\ su(3)\atop {\rm or}\ \widehat{su}(3) & SU(4)  \ {\rm or }\ su(4)  \atop {\rm or}\  \widehat{su}(4) &
   G \ {\rm or }\ \gg   \atop  { \hbox{\tiny other simple Lie} \atop  \hbox{\tiny group or Lie algebra}}&  
  \widehat{\gg} \atop{\hbox{\tiny other affine Lie algebra}} &\Gamma  \atop  {\hbox{\tiny finite simple}\atop\hbox{\tiny group of Lie type}}&\Gamma  \atop  \hbox{\tiny other finite group}
 \\[4pt]
\hline \hline
m_2=\bar m_2  & \cm & \cm & \cm & \cm &\cm &\cm \\
\hline
m_1=\bar m_1 &  \cm & \cm &\cm & \cm  &\cm ? &{\rm X}  \\
\hline
m_0=\bar m_0 & \cm & \cm\, ? & {\rm X} & {\rm X} &  {\rm X} &{\rm X} \\
\hline
m_r=\bm_r\ \forall r
\atop \Leftrightarrow\quad \mathfrak P & \cm & {\rm X} & {\rm X} & {\rm X} &  {\rm X} &{\rm X} \\
\hline\hline
\end{array}
$
\end{center}
\caption{{$\cm$ means that the property is true and proven ; X that it is not true in general and there are counter-examples ;
$\cm\,?$ that the property has been checked 
in many cases (see text) but that a general proof is still missing.}\label{results-ext}}
\end{table}

\subsection{Comments, remarks, examples and counter-examples}
 \begin{itemize}
 \item The equality $m_2=\bar m_2$ is the easiest to interpret and to prove. More explicitly it asserts that
 \be\label{CZ2}\sum_\nu (N_{\lambda\mu}^\nu)^2=\sum_{\nu'} (N_{\lambda\bar\mu}^{\nu'})^2\,. \ee
Proof. The number of invariants $N_{\lambda\mu \bar\lambda\bar\mu}^{0}$ in 
${\lambda\otimes \mu\otimes\bar\lambda\otimes\bar\mu}$ may be written as
\bea\nonumber N_{\lambda\mu \bar\lambda\bar\mu}^{0}&\eq={\rm {(i)}}&
\sum_{\nu,\nu'} N_{\lambda\mu}^\nu N_{\bar\lambda\bar\mu}^{\nu'}N_{\nu\nu'}^0
\eq={\rm {(ii)}}\sum_{\nu,\nu'} N_{\lambda\mu}^\nu N_{\bar\lambda\bar\mu}^{\nu'}\delta_{\nu'\bar\nu}
\eq={\rm {(iii)}}\sum_\nu N_{\lambda\mu}^\nu N_{\bar\lambda\bar\mu}^{\bar\nu}= \sum_\nu (N_{\lambda\mu}^\nu)^2\\
&\eq={\rm {(iv)}}&N_{\lambda\bar\mu \bar\lambda\mu}^{0}=\sum_\nu N_{\lambda\bar\mu}^\nu N_{\bar\lambda\mu}^{\bar\nu}=\sum_\nu (N_{\lambda\bar\mu}^\nu)^2 \eea
where we have made use of (i) associativity of the tensor or fusion product, 
(ii) $N_{\nu\nu'}^0=\delta_{\nu'\bar\nu}$,  (iii) invariance under conjugation $N_{\bar\lambda\bar\mu}^{\bar\nu}=N_{\lambda\mu}^\nu$, and (iv) commutativity $N_{\lambda\mu \bar\lambda\bar\mu}^{0}=N_{\lambda\bar\mu \bar\lambda\mu}^{0}$.

Graphically, this may be represented as in Fig. \ref{su-crossing}.  
In physical terms, and in the context of particle physics,
 it expresses the fact that the numbers of independent amplitudes in the ``$s$ channel"
$\lambda \otimes \mu \rightarrow \lambda \otimes \mu$ and in the ``crossed $u$ channel"
$\lambda \otimes \bar\mu \rightarrow \lambda \otimes \bar \mu$ are the same.
 \begin{figure}[ht]
\centering{\includegraphics[width=8cm]{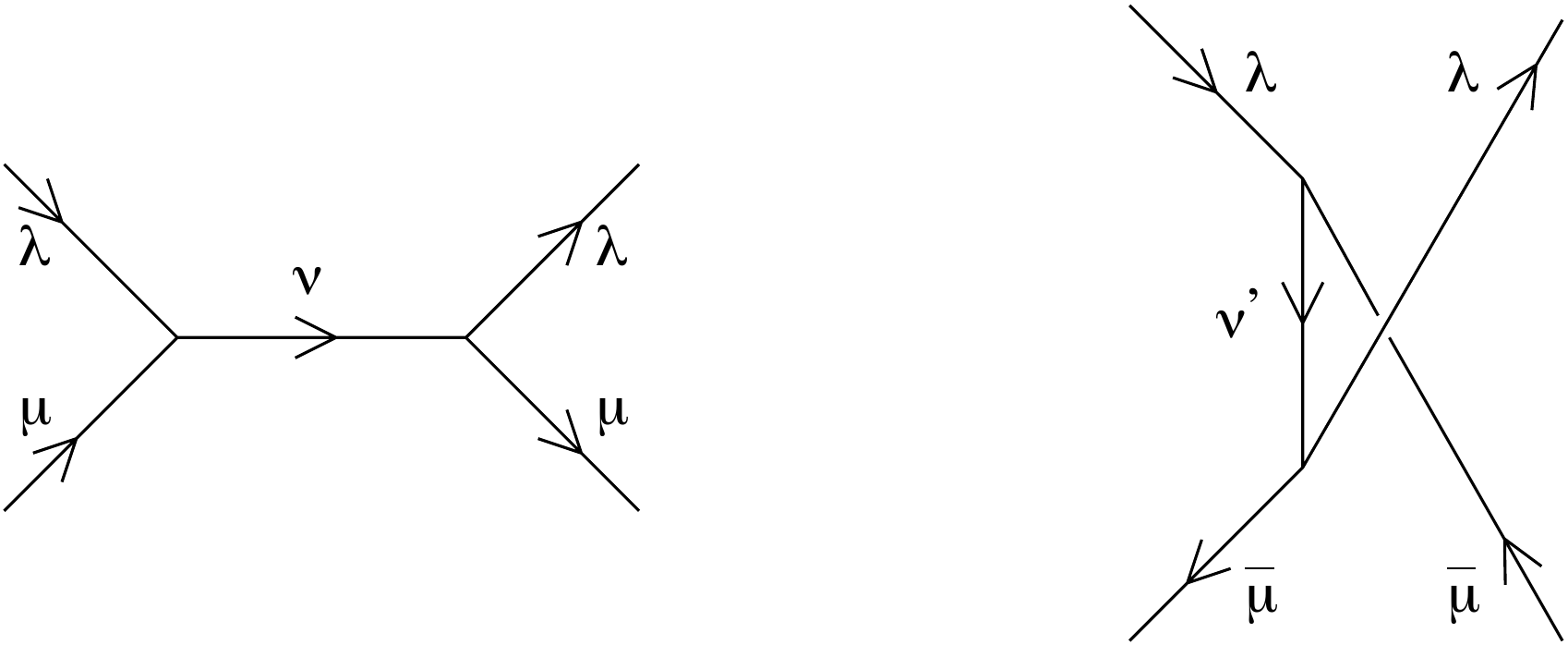}}\\ 
\caption{\label{su-crossing}Graphical representation of $m_2=\bar m_2$.
Each  $\lambda\mu\nu$ vertex carries the multiplicity $N_{\lambda\mu}^\nu$, and likewise for  $\lambda\bar\mu\nu'$
on the right. Sums over $\nu$, resp. $\nu'$ are equal.}
\end{figure} 
\item In contrast, the equality $m_1=\bar m_1$ is neither natural nor general. While it is valid for all 
simple Lie algebras, either finite dimensional or affine, (see the discussion and elements of proofs in the next section), it is
known not to be true for general finite groups. Counter-examples are provided by some finite subgroups of SU(3),
see below in sect.\  \ref{finitegps}, and also \cite{RCJBZ2011}, and the detailed discussion in  \cite{RCJBZ2013}.
\item Even more elusive and exceptional is the equality $m_0=\bar m_0$, which happens to be true in SU(3) or for
the affine algebra $\widehat{su}(3)$, as a particular case of the more general property $\mathfrak P$ that they satisfy. 
Curiously we have found evidence (but no proof yet) that it also holds true for SU(4) and $\widehat{su}(4)$ 
(this was tested in $\widehat{su}(4)_k$ up to level $k=15$), but it fails 
in general for higher rank SU($N$) or $\widehat{su}(N)$.  
\item Finally the equality  $m_r=\bar m_r$ for all $r$, or equivalently property $\mathfrak P$, is satisfied in SU(3) \cite{RCJBZ2014}
and in  $\widehat{su}(3)$ at all levels \cite{RCJBZ2016}.
 \end{itemize}
 
 Example 1. In SU(3), for the ten-dimensional representations,  %From Lie
% X[2,1] * X[2,1] = 1X[4,2] +1X[5,0] +1X[2,3] +2X[3,1] +1X[0,4] +2X[1,2] +1X[2,0] +1X[0,1]
 %X[2,1] * X[1,2]  = 1X[3,3] +1X[4,1] +1X[1,4] +2X[2,2] +1X[3,0] +1X[0,3] +2X[1,1] +1X[0,0]
 \bea\nonumber
  (2,1) \otimes (2,1) &=& 1(4,2) +1(5,0) +1(2,3) +2(3,1) +1(0,4) +2(1,2) +1(2,0) +1(0,1)\\
  (2,1) \otimes (1,2) &=& 1(3,3) +1(4,1) +1(1,4) +2(2,2) +1(3,0) +1(0,3) +2(1,1) +1(0,0)
 \eea
 on which we do observe all the above properties: $m_2=\bm_2=14$, $m_1=\bm_1=10$, $m_0=\bm_0=8$
 and the multisets of multiplicities are both $\{1,1,1,1,1,1,2,2\}$, or in short, $\{1^6 2^2\}$
(where we note the number $n$ of occurrences of multiplicity $m$ by $m^n$). 
 
 Example 2. In SU(4), 
 with $\lambda=\mu=(1,2,2)$, we find for the multiplicities $N_{\lambda\mu}^\nu$
 the multiset 
% 1X[2,4,4] +1X[3,2,5] +1X[2,5,2] +1X[0,5,4] +1X[4,0,6] +2X[3,3,3] + 1X[2,6,0] +2X[1,3,5] +1X[0,6,2] +2X[4,1,4] +2X[3,4,1] +2X[2,1,6] + 4X[1,4,3] +1X[0,7,0] +3X[4,2,2] +5X[2,2,4] +4X[1,5,1] +1X[0,2,6] + 1X[5,0,3] +1X[4,3,0] +3X[3,0,5] +7X[2,3,2] +1X[1,0,7] +3X[0,3,4] + 2X[5,1,1] +6X[3,1,3] +3X[2,4,0] +4X[1,1,5] +4X[0,4,2] +6X[3,2,1] + 7X[1,2,3] +2X[0,5,0] +1X[6,0,0] +3X[4,0,2] +4X[2,0,4] +6X[1,3,1] + 1X[0,0,6] +3X[4,1,0] +7X[2,1,2] +3X[0,1,4] +4X[2,2,0] +4X[0,2,2] + 3X[3,0,1] +3X[1,0,3] +2X[0,3,0] +4X[1,1,1] +1X[2,0,0] +1X[0,0,2] +1X[0,1,0]
$\{1^{17} 2^8 3^9 4^8 5^1 6^3 7^3\}$ 
%(where we note the number $n$ of occurrences of multiplicity $m$ by $m^n$)
%
%1X[3,4,3] +1X[4,2,4] +1X[3,5,1] +1X[1,5,3] +1X[5,0,5] +2X[4,3,2] + 2X[2,3,4] +1X[1,6,1] +2X[5,1,3] +1X[4,4,0] +2X[3,1,5] +4X[2,4,2] + 1X[0,4,4] +2X[5,2,1] +5X[3,2,3] +2X[2,5,0] +2X[1,2,5] +2X[0,5,2] +1X[6,0,2] +3X[4,0,4] +5X[3,3,1] +1X[2,0,6] +5X[1,3,3] +1X[0,6,0] + 1X[6,1,0] +5X[4,1,2] +5X[2,1,4] +5X[1,4,1] +1X[0,1,6] +3X[4,2,0] + 8X[2,2,2] +3X[0,2,4] +2X[5,0,1] +5X[3,0,3] +4X[2,3,0] +2X[1,0,5] + 4X[0,3,2] +6X[3,1,1] +6X[1,1,3] +2X[0,4,0] +6X[1,2,1] +1X[4,0,0] + 5X[2,0,2] +1X[0,0,4] +3X[2,1,0] +3X[0,1,2] +2X[0,2,0] +3X[1,0,1] +1X[0,0,0]
while for those for $\lambda\otimes \bmu$ it is
$\{1^{16} 2^{12} 3^6 4^3 5^8 6^3 8^1\}$, whence $m_2=\bm_2=538$, $m_1=\bm_1=136$, 
$m_0=\bm_0=49$ but the multisets are clearly different.
%Mathematica 
%1 17 + 2 8 + 3 9 + 4 8 + 5 1 + 6 3 + 7 3
%1 16 + 2 12 + 3 6 + 4 3 + 5 8 + 6 3 + 8 1
%1 17 + 2^2 8 + 3^2 9 + 4^2 8 + 5^2 1 + 6^2 3 + 7^2 3
%1^2 16 + 2^2 12 + 3^2 6 + 4^2 3 + 5^2 8 + 6^2 3 + 8^2 1
 
 Example 3. In SU(5), for $\lambda=(1,1,1,0)$, $\mu=(1,1,0,1)$, we find that 
 the list of $N_{\lambda\mu}^\nu$ reads $\{1^{12} 2^6 3^3 4^3\}$ 
 %1X[2,2,1,1] +1X[3,0,2,1] +1X[0,3,1,1] +1X[3,1,0,2] +1X[2,3,0,0] + 2X[1,1,2,1] +2X[3,1,1,0] +2X[1,2,0,2] +1X[0,4,0,0] +2X[2,0,1,2] + 4X[1,2,1,0] +1X[0,0,3,1] +1X[4,0,0,1] +3X[2,0,2,0] +2X[0,1,1,2] + 4X[2,1,0,1] +3X[0,1,2,0] +1X[1,0,0,3] +3X[0,2,0,1] +4X[1,0,1,1] + 1X[3,0,0,0] +2X[1,1,0,0] +1X[0,0,0,2] +1X[0,0,1,0]
  while that of $N_{\lambda\bmu}^{\nu'}$ reads $\{1^{15}  2^3 3^4 4^3\}$.
% 1X[2,1,2,1] +1X[2,2,0,2] +1X[0,2,2,1] +1X[3,0,1,2] +1X[2,2,1,0] +1X[1,0,3,1] +1X[0,3,0,2] +1X[3,0,2,0] +3X[1,1,1,2] +1X[0,3,1,0] + 2X[3,1,0,1] +3X[1,1,2,0] +1X[2,0,0,3] +4X[1,2,0,1] +1X[0,0,2,2] + 4X[2,0,1,1] +1X[0,1,0,3] +1X[0,0,3,0] +1X[4,0,0,0] +4X[0,1,1,1] +3X[2,1,0,0] +2X[1,0,0,2] +2X[0,2,0,0] +3X[1,0,1,0] +1X[0,0,0,1]
 We check that $m_2=\bm_2=111$ and $m_1=\bm_1=45$ but $m_0=24\ne \bm_0=25$. 
 
 Example 4. In SO(10) (Lie algebra $D_5$), with 
 \ommit{$\lambda=(1, 1, 1, 1, 0)$  and $\mu=(1, 1, 0, 1, 0)$,
 the two multisets are respectively $\{1^{27} 2^{18} 3^6 4^{11} 5^{10} 6^4 7^3 8^4 9^6 10^8 12^3 13^2 14^3 15^2 17^1 21^1 22^1 26^1\}$ and  \\ $\{1^{23} 2^{20} 3^{12}
 4^8 5^8 6^7 7^4 8^5 9^3 10^3 11^3 12^5 14^3  15^2 17^1 21^2 22^2\}$
 from which we check that $m_2=\bm_2=6110$, $m_1=\bm_1=606$, $m_0=\bm_0=111$ while the two multisets are
 obviously different.}
$\lambda=\mu=(1,1,0,1,0)$, the two multisets are respectively 
$\{1^{17}  2^{10}  3^3  4^8  5^6  6^3  7^2  8^2   12^1  \}$
and $\{1^{15}  2^{11}  3^8  4^7  5^2 6^3  7^1  8^2  9^2  10^1\}$ 
   from which we check that $m_2=\bm_2=840$, $m_1=\bm_1=168$, $m_0=\bm_0=52$ while the two multisets are
manifestly different.
 
 Example 5. In SO(10) (Lie algebra $D_5$), with $\lambda=\mu=(1, 1, 1, 1, 0)$,
we find  $m_1=\bm_1=4456$  et $m_2=\bm_2=184216$ but 
$m_0=240$ and $\bm_0=243$, hence a counter-example to the property of Prop.\  5.

Example 6. In $E_6$, likewise, we may find pairs of $\lambda,\, \mu$ which violate Prop.\  4 and 5. 
Take $\lambda=\mu= %[1,1,1,0,0,0]$, $\bmu=[0,1,0,0,1,1]$ (in LiE's conventions, \ie in ours, $\lambda=\mu=
(1, 1, 0, 0, 0;1) ; \bmu = (0, 0, 0, 1, 1; 1)$\footnote{We use the common convention that
the component of the vertex located on the short branch of the Dynkin diagram is written at the end}; 
we find $m_1=\bm_1=947$, 
$m_2=\bm_2=14163$ but $m_0=119$, $\bm_0=123$. (Incidentally, the reducible representation encoded
by $\lambda\otimes \mu$ in that case has dimension $63631071504 = 252252^2$. )

 %%%%%%%%%%%%%%%%%%%%
  
 \subsection{Related properties of the modular $S$-matrix}
 In the case of an affine algebra $\widehat{\gg}_k$, it is well known that the fusion coefficients
 are given by Verlinde formula\,\cite{Verlinde}
 \be\label{verlinde} N_{\lambda\mu}^\nu=\sum_\kappa \frac{S_{\lambda\kappa}S_{\mu\kappa}S_{\nu\kappa}^*}{S_{0\kappa}}\,.\ee
 
 {\bf Proposition 6} \cite{RCJBZ2011}{\sl\  For the affine algebra $\widehat{\gg}_k$ at finite level $k$,
$\Sigma(\kappa):=\sum_\nu S_{\kappa\nu}$ vanishes if the irrep $\kappa$ is either of complex or of quaternionic type.}

For $\kappa$ complex, $\kappa\ne \bar\kappa$, this implies immediately Proposition 2, since, 
using the fact that $S_{\mu\bar\kappa}=S_{\bmu\kappa}=S_{\mu\kappa}^*$, 
\be\sum_{\nu} N_{\lambda\mu}^\nu=\sum_{\kappa}
\frac{S_{\lambda\kappa}S_{\mu\kappa}\sum_\nu S_{\nu\kappa}^*}{S_{0\kappa}}=\sum_{\nu,\kappa=\bar\kappa}
\frac{S_{\lambda\kappa}S_{\mu\kappa}S_{\nu\kappa}^*}{S_{0\kappa}}=\sum_{\nu,\kappa=\bar\kappa}
\frac{S_{\lambda\kappa}S_{\bmu\kappa}S_{\nu\kappa}^*}{S_{0\kappa}}=\sum_{\nu} N_{\lambda\bmu}^\nu\,.\ee
 But conversely, as
shown %below 
in \cite{RCJBZ2011} by a fairly simple argument, Prop.\  2 implies that $\Sigma(\kappa)=0$ if $\kappa\ne\bar\kappa$. 

The fact that the sum $\Sigma(\kappa)$ also vanishes for $\kappa$, a representation of quaternionic type, 
though of no direct relevance for the present discussion, is also a curious observation and was proved in 
\cite{RCJBZ2011} as the result of a case by case analysis.

%%%%%%%%%%%%%%%%%%%%%%

\subsection{The case of finite groups}
\label{finitegps}
\begin{figure}[ht]
\centering{\includegraphics[width=14cm]{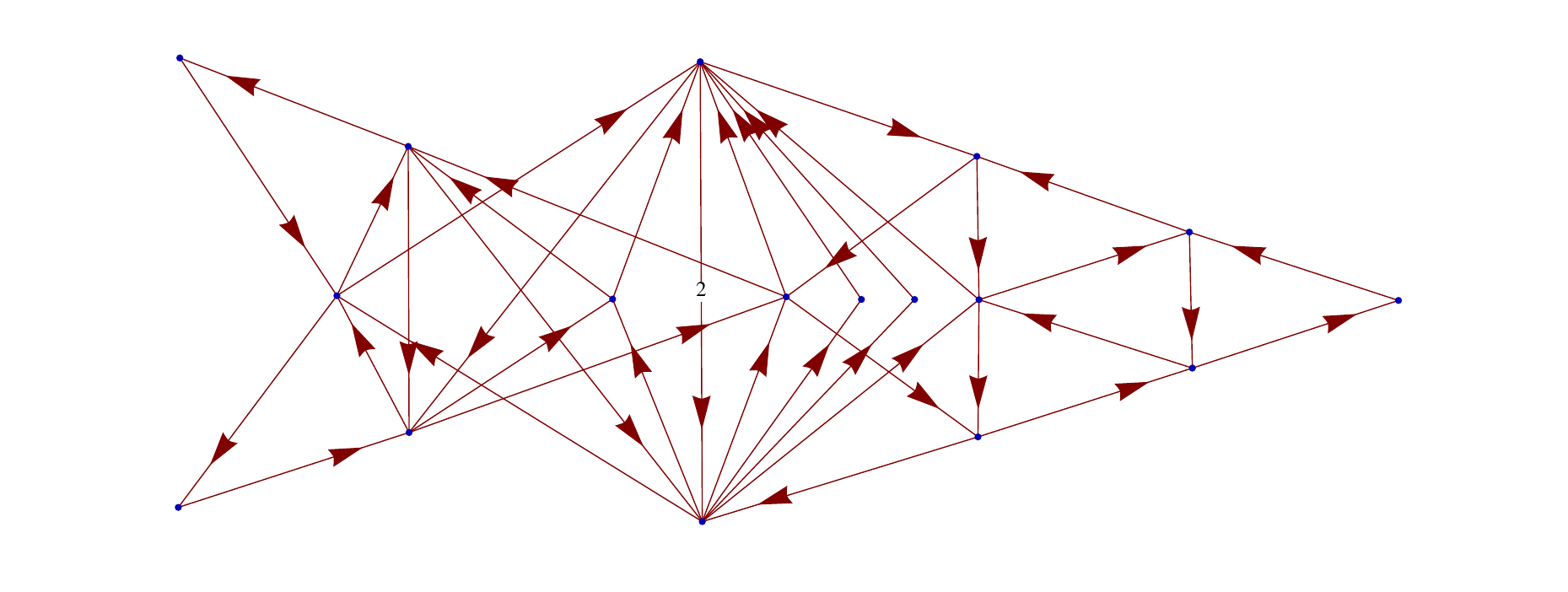}}\\
\caption{\label{Sigma1080}Tensor product graph for the subgroup $\Sigma(3\times 360)$ of SU(3).
Note: the middle vertical edge carries a multiplicity 2.}
\end{figure}

%Example 7. 
Finite groups (admitting complex representations) do not generally satisfy Prop.\ 2.
Consider for example the finite subgroup $\Gamma=\Sigma(3\times 360)$ of SU(3)\,\cite{YY,FFK}.
 A simple way to show that  the equality of $m_1$ and $\bm_1$ 
is not satisfied is to draw the oriented graph whose vertices are the irreps of $\Gamma$ and 
whose adjacency matrix is the matrix $N_{f\mu}^\nu$, where $f$ denotes 
one of the 3-dimensional irreducible representations, see Fig. 
\ref{Sigma1080}.
In Fig. \ref{Sigma1080}, pairs of complex conjugate representations are images under a reflection through 
the horizontal axis. 
 The sum $\sum_\nu N_{f\mu}^\nu$ counts the number of oriented edges exiting vertex $\mu$.
It is clear that the sums relative to $\mu=$ the outmost  upper and lower vertices are different. \\
Ultimately, we found among subgroups of SU(3) the following counter-examples \cite{RCJBZ2011} to Prop.\ 2:
$\Sigma(3\times 72)$, $\Sigma(3\times 360)$, and the subgroups  of the type $F_{3m}=\mathbb{Z}_m\rtimes\mathbb{Z}_3$, where $m$ should be a prime of the type $6p+1$. % \cite{RCJBZ2011}.} 

Discussion. 
Could the validity of Prop.\ 2 be related to the {\it modularity} of the tensor (or fusion) category, 
which holds true for Lie groups and affine algebras, but not generally for finite groups? 
In \cite{RCJBZ2013} we explored that possibility by constructing the Drinfeld doubles 
of subgroups of SU(2) and SU(3). While tensor product in Drinfeld doubles is  known to be
modular, we found again many counter-examples to Prop.\ 2, in particular for the double of the same group
 $\Sigma(3\times 360)$. We conclude that the property encapsulated in Prop.\  2 is not a 
 modular property but rather seems to be a Lie theory property. See below in sect.\  2 a remark on the role of the 
 Weyl group in the proof. 

\bigskip

The validity of Prop.\  2 is not directly related, either, to the simplicity of the group considered; indeed, the Mathieu groups $M_{11}$, $M_{12}$, $M_{21}$
$M_{22}$, $M_{23}$, $M_{24}$ are simple finite groups, but Prop.\  2 is only valid for $M_{12}$ and $M_{21}$ (the latter, although simple, does not appear in the list of sporadic simple groups because it is isomorphic with $PSL(3,4)$). 

\ommit{A modifier :
However we also considered those Chevalley groups that admit complex representations -- otherwise Prop.\  2 is trivially verified. For small values of $n\geq 1$ and $q$ (a power of a prime) 
we looked at examples from the families $A_n(q)=SL(n+1,q)$,  $B_n(q)=O(2n+1,q)$, $C_n(q)=Sp(2n,q)$, $D_n(q)=\Omega^{+}(2n,q)$, $G_2(q)$, 
and also from the families called $2A_n(q)=SU(n+1,q)$,  $2B_n(q)=Sz(q)$ (Suzuki), $2D_n(q)=\Omega^{-}(2n,q)$, $2G_2(q)$ (Ree), $3D_4(q)$, with the notation of MAGMA \cite{},
 and did not find a single counter-example to Prop.\  2 when the finite group of Lie type under study was simple.
 We could not explicitly study examples from the families $F_4(q)$, $E_6(q)$, $E_7(q)$, $E_8(q)$, or $2F_4(q)$ (Ree), $2E_6(q)$,
because of the size of their character table. Nevertheless, the obtained results suggest (on en fait une conjecture ?) that Prop.\  2 is valid for all simple groups of Lie type.
The simplicity argument seems here to be important: for example, in the family $2A_n(q)=SU(n+1,q)$, we checked that Prop.\ 2 holds for $2A_1(q)$, $q=2,3,4,5,7$ (it holds trivially for $q=2,4,5$), 
$2A_2(3)$, $2A_2(4)$, $2A_2(7)$, $2A_3(2)$, $2A_4(2)$, but does not hold for $2A_2(2)$, $2A_2(5)$, $2A_3(3)$, however the last three are not simple. 
For finite groups of Lie type, this simplicity condition seems to be sufficient but not necessary since we also found several non-simple groups that admit complex representations and verify Prop.\  2.
The ``biggest'' simple group of Lie type that we considered (and obeying Prop.\ 2)  is $G_2(5)$, with  $5859000000$ elements, $44$ conjugacy classes (or irreps), and only four complex irreps.}

We also considered those Chevalley groups that admit complex representations -- otherwise Prop.\ \ ~2 would be trivially verified. 
For small values of $n\geq 1$ and $q$ (a power of a prime) 
we looked at examples from the families $A_n(q)=SL(n+1,q)$,  $B_n(q)=O(2n+1,q)$, $C_n(q)=Sp(2n,q)$, $D_n(q)=\Omega^{+}(2n,q)$, $G_2(q)$, 
and also from the families called $2A_n(q)=SU(n+1,q)$,  $2B_n(q)=Sz(q)$ (Suzuki), $2D_n(q)=\Omega^{-}(2n,q)$, $2G_2(q)$ (Ree), $3D_4(q)$, with the notations used in MAGMA \cite{Magma}.
We could not explicitly study examples from the families $F_4(q)$, $E_6(q)$, $E_7(q)$, $E_8(q)$, or $2F_4(q)$ (Ree), $2E_6(q)$, because of the size of their character table. 
The largest simple group of Lie type that we considered (and obeying Prop.\ 2)  was $G_2(5)$, with  $5859000000$ elements, $44$ conjugacy classes (or irreps), and only four complex irreps.
All together we tested about 70 Chevalley groups, 33 of them being simple,  and 37 had complex irreps, so that testing the sum rule (Prop.\ 2) for them was meaningful.
Among those 37 groups with complex irreps, 21 were simple and the sum rule was obeyed by all of them; among the $37-21=16$ non-simple groups %(37-21) 
with complex irreps, we %with 
found 4 cases for which the sum rule fails. 
% ton addition suggŽrŽe, OK ?
In all cases where this sum rule failed for a non-simple Chevalley group, it turned out to hold for the corresponding projective group (a simple quotient of the latter):  for instance the rule fails for the non-simple group $A_2(7) = SL(3,7)$ but it holds for the simple group $PSL(3,7)$ (and also holds for the non-isomorphic simple group $2A_2(7) = SU(3,7)=PSU(3,7)$).
Although the obtained results may not be statistically significant they seem to indicate that Prop.\  2 is valid for simple groups of Lie type. 
We did not try to prove this property but if it happens to be true, one may expect, for finite groups of Lie type, that the Weyl group could play a role in the proof,  like in the case of simple Lie groups (see below).

 %%%%%%%%%%%%%%%%%%
\section{A sketch of proofs}

The proof of Prop.\  1 has been given above. We shall content ourselves with sketches of proofs for the other 
propositions. 

The proof of Prop.\  2 may be split in two steps.\\
 {\bf Lemma 1.} {\sl\  Prop.\  2 holds for $\lambda=\omega_p$, a fundamental weight.}\\
 {\bf Lemma 2.} {\sl\  Prop.\  2 holds for any product of the fundamental representations.}\\
\begin{proof}
The first lemma was established   for the $A_r,\ D_{r \, \text{odd}},\ E_6$ 
simple Lie algebras (the others do not have complex irreps)
making use of the Racah-Speiser formula, or of its affine extension. 
The latter expresses the tensor coefficient $N_{\lambda\mu}^\nu$  as a weighted sum over 
suitable elements of the (classical or affine) 
Weyl group, 
see  \cite{RCJBZ2011} for details. Restricting $\lambda$ to be a fundamental
weight makes the discussion amenable to a fairly simple  
analysis of a finite number of cases.

\ommit{states that \dots
\be\label{rasp}
N_{\lambda\mu}^{\nu}=\sum_{\lambda'\in [\lambda]}\sum_{w\in W \atop  w[\lambda'+\mu + \rho]-\rho\in \mathfrak P_+} {\rm sign}(w)\,
\delta_{\nu, w[\lambda'+\mu + \rho]-\rho}
\ee}

The second lemma follows simply from the associativity and commutativity of the tensor or fusion product,
using the outcome of Lemma 1:
$$ \sum_{\nu}\left(N_{\omega_p}\right)_\mu^{\  \nu}=\sum_{\nu}\left(N_{\bar \omega_p}\right)_\mu^{\  \nu}\ .$$
This, together with the commutativity of the $N$ matrices, 
 entails that for any monomial $N_{\omega_{j_1}}\cdots N_{\omega_{j_q}}$
\bea \sum_{\nu}\left(N_{\omega_{j_1}}\cdots N_{\omega_{j_q}}\right)_\mu^{\  \nu} 
&=&\sum_{\nu'}\left(N_{\omega_{j_1}}\cdots N_{\omega_{j_{q-1}}}\right)_\mu^{\  \nu'}\sum_{\nu }\left( N_{\omega_{j_q}}\right)_{\nu'}^{\ \nu}\nonumber \\
&\buildrel {\rm (Lemma\ 1)}\over =&\sum_{\nu'}\left(N_{\omega_{j_1}}\cdots N_{\omega_{j_{q-1}}}\right)_\mu^{\  \nu'}\sum_{\nu }\left( N_{\bar\omega_{j_q}}\right)_{\nu'}^{\ \nu}\nonumber  \\
&=&\sum_{\nu}\left(N_{\bar\omega_{j_q}}N_{\omega_{j_1}}\cdots N_{\omega_{j_{q-1}}}\right)_\mu^{\  \nu}=\cdots\nonumber  \\
&=&\sum_{\nu}\left(N_{\bar\omega_{j_1}}\cdots N_{\bar\omega_{j_q}}\right)_\mu^{\  \nu}
\eea

This completes the proof of the two lemmas. As any $N_\lambda$ is a polynomial in the commuting $N_{\omega_p}$, 
$p=1,\cdots,r$, 
$N_\lambda=P_{\lambda}(N_{\omega_1},\cdots,N_{\omega_{r}})$
and  $N_{\bar\lambda}=P_{\lambda}(N_{\bar\omega_{1}},\cdots,N_{\bar\omega_{r}})$, this also
establishes  Prop.\  2.
\end{proof}

The salient feature of this approach is the crucial role played by the (classical or affine) Weyl group.

Prop. 3 and 4, which deal with the explicit case of the classical or affine $su(3)$ algebra, 
have been established through a detailed and laborious analysis which will not be repeated here. 
We only mention  that a variety of graphical representations of the determination of the $N_{\lambda\mu}^\nu$
coefficients has been used.
We refer the reader to \cite{RCJBZ2014} and \cite{RCJBZ2016} for details.

%%%%%%%%%%%%%%%%%%%%
\section{Conclusion}
In this letter, we have reviewed some recent results on conjugation properties of tensor product (or fusion) 
multiplicities. Although quite simple to state, it appears that these results were not previously known, and
that some are fairly difficult to prove. In particular we feel that our proofs of Propositions 2, 3 and 4 lack elegance and
may miss some essential concept. Hopefully some inspired reader will come with  new insight in these matters.
%%%%%%%%%%%%%%%%%%%%%%%%%%%%%%%%%%%%%%%%%%%%%%%%%%%%%%%%%%%

\newpage

\end{document}